\documentclass[showpacs, preprintnumbers, amsmath, amssymb]{revtex4}

\usepackage{dcolumn}% Align table columns on decimal point
\usepackage{bm}% bold math

\usepackage{epsfig}

\begin{document}

\title{Title}

\author{M. O. Hase and H. L. Casa Grande}

\affiliation{Escola de Artes, Ci\^encias e Humanidades, Universidade de S\~ao Paulo\\ Avenida Arlindo B\'ettio 1000, 03828-000 S\~{a}o Paulo, Brazil}
\email{mhase@usp.br}

\title{Carrying capacity in growing networks}

\begin{abstract}
In this work, a growing network model that can generate a random network with finite degree in infinite time is studied. The dynamics are governed by a rule where the degree increases under a scheme similar to the Malthus-Verhulst model in the context of population growth. The degree distribution is analysed in both stationary and time-dependent regimes through some exact results and simulations, and a scaling behaviour is found in asymptotically large time. For finite times, the time-dependent degree distribution displays an accumulation of hubs as a result of competition between attractive and repulsive terms in linking probability.
\end{abstract}


\maketitle

 %%%%%%%%%%%%%%%%%%%%%%%%%%%%%%%%%%%%%%%%%%%%%%%%%%%%%%%%%%%%%%%%%%%%%%%%%%%%%%%%%%%%%%%%%%%%%%%%%%%%%%%%%%%%%%
 %%%%%%%%%%%%%%%%%%%%%%%%%%%%%%%%%%%%%%%%%%%%%%%%%%%%%%%%%%%%%%%%%%%%%%%%%%%%%%%%%%%%%%%%%%%%%%%%%%%%%%%%%%%%%%
 %%%%%%%%%%%%%%%%%%%%%%%%%%%%%%%%%%%%%%%%%%%%%%%%%%%%%%%%%%%%%%%%%%%%%%%%%%%%%%%%%%%%%%%%%%%%%%%%%%%%%%%%%%%%%%

\section{Introduction}

Several complex systems, such as the World Wide Web, social networks of acquaintance, metabolic networks, and many others, are known to exhibit a heterogeneous behaviour in their degree distributions, which is usually a power-law function \cite{AB02, N03}. Furthermore, they are not static graphs, but their size (number of vertices and degrees) changes with time. A simple model for growing networks proposed in the late 1990s by Barab\'asi and Albert \cite{BA99} (BA) can generate a power-law degree distribution, and two ingredients were recognized to be sufficient to generate such behaviour. Firstly, the model is a non-static growing network, which means that vertices are continuously added to the graph. Secondly, the new incoming vertices attach to old ones ruled by (linear) preferential linking \cite{BA99, DMS00}; this means that highly connected vertices are likely to be linked by the new ones with a probability proportional to its degree. It is worthwhile remarking that in some cases, when this linearity is broken, the model may fail to display the fat-tailed distribution \cite{KR01, KRL00}.

The BA model has important extensions where many ideas can be tested. To cite some of them, the \textquotedblleft fitness\textquotedblright of a vertex is introduced in Bianconi-Barab\'asi model \cite{BB01}. The BA model is also widely used to investigate network properties such as resilience \cite{AJB00, CEbAH00} or to study critical phenomena by defining statistical models on it \cite{DGM08}. In the same spirit, the present work will explore the following question: the BA model makes highly connected vertices attractive, but the model does not impose any limit for the degree, which is unbounded and increases with time. However, in some known situations (like in the \textit{El Farol} bar problem \cite{A94}), \textquotedblleft crowded\textquotedblright vertices are not attractive. The competition between preferential linking and unattractiveness of highly connected nodes can be inserted in a simple statistical model. There are some previous works where a cutoff is introduced in order to limit the degree or studies concerning the size of the networks \cite{ASBS00, DMS01, MBSA02, BP-SV04}. In the literature, there are others means to avoid the indefinite increasing of a degree by considering finite systems \cite{KR02} or ageing effect on vertices. The ageing effect is introduced by making a vertex inactive to receive more links with some probability at each time step \cite{ASBS00}, or simply decreasing the probability of receiving a connection with time \cite{DM00}. In the latter prescription, although the maximum degree $k_{\textnormal{max}}$ of a vertex still increases with time $t$, the ratio $k_{\textnormal{max}}/t\rightarrow 0$ as $t\rightarrow\infty$.

Nevertheless, in this work, the form by which an indefinite growth of degree is avoided is inspired by an old idea that traces back to some classical models of population dynamics. The simplest paradigm in this context seems to be the Malthusian model \cite{M98}, which assumes that the growth of a population is proportional to its actual size. This model predicts an exponential growth for the number of species, but does not take into account the mechanisms that prevent arbitrary increasing. A simple idea in this direction points to Verhulst \cite{V58}, where the idea of carrying capacity, which is the maximum population number, is introduced. In the Malthus-Verhulst model, the population increases as a sigmoid function, which means that the growth rate is smoothly decreased, and not stopped by a cutoff. In this sense, the population \textquotedblleft is aware \textquotedblright of its maximum capacity from the very beginning. The main idea of this work is to model the growth of a degree in a scheme analogous to the Malthus-Verhulst model, and show that this proposal leads to a behaviour not predicted by many of previous models (but found in  \cite{KR02}), which is the accumulation of hubs.

The layout of this paper is as follows: the main problem and the model are introduced in sections 2 and 3, and its static properties are examined in the following section. The time-dependent analysis is performed in section 5, where an unexpected behaviour of the degree distribution is discussed, and a toy model is introduced in section 6 to complement the analysis. General observations and conclusions are presented in the section 7.

 %%%%%%%%%%%%%%%%%%%%%%%%%%%%%%%%%%%%%%%%%%%%%%%%%%%%%%%%%%%%%%%%%%%%%%%%%%%%%%%%%%%%%%%%%%%%%%%%%%%%%%%%%%%%%%
 %%%%%%%%%%%%%%%%%%%%%%%%%%%%%%%%%%%%%%%%%%%%%%%%%%%%%%%%%%%%%%%%%%%%%%%%%%%%%%%%%%%%%%%%%%%%%%%%%%%%%%%%%%%%%%
 %%%%%%%%%%%%%%%%%%%%%%%%%%%%%%%%%%%%%%%%%%%%%%%%%%%%%%%%%%%%%%%%%%%%%%%%%%%%%%%%%%%%%%%%%%%%%%%%%%%%%%%%%%%%%%

\section{Formulation of the problem}

The network considered in this work is a growing one that will be described by a discrete time master equation following  \cite{DM03}, as shown later. The initial condition, conveniently taken at time $t=2$, consists of two vertices, denoted by $s=1$ and $s=2$, linked to each other by two edges. At each unitary time step, a new vertex is added to the graph and links with one of the previously existent vertex with conditional probability $\Pi$, which will be assumed to have the same form for any vertex of the graph. Note that the variable $s$ labels the vertices and coincides with the time when it joined the network. A key quantity to perform the analysis of this system is $p(k, s, t)$, which is the probability that a vertex $s$ has $k$ connections at time $t$. The discrete time master equation is then
\begin{align}
p(k, s, t+1) = \Pi(k-1, t)p(k-1, s, t) + \big[1-\Pi(k, t)\big]p(k, s, t)\,.
\label{me}
\end{align}
This equation can be easily generalised for a growing network that allows more links per time step, but this is the simplest setup. Note that the total number of degrees of the graph is $2t$ at time $t$. It is clear that one has initially $p(k, s=1, t=2)=p(k, s=2, t=2)=\delta_{k,2}$, where $\delta_{x,y}$ stands for Kronecker delta (as usual, $\delta_{x,y}=1$ if $x=y$ and $\delta_{x,y}=0$ otherwise), and $p(k, s=t, t)=\delta_{k,1}$ stands for the boundary condition indicating the addition of a single connection from the newly added vertex.

In BA model \cite{BA99}, the new incoming vertex connects with an old one with a probability proportional to the number of the degree of the latter, which implies $\Pi_{BA}(k, t)\propto k$. This choice makes popular vertices more likely to be connected (\textquotedblleft rich gets richer\textquotedblright), displays a degree distribution that follows a power-law behaviour \cite{BA99, KRL00, DMS00}, and there is no upper bound for the degree that a vertex is allowed to have. In order to introduce a mechanism that prevents an indefinite growth of the degree, let
\begin{align}
\Pi = \Pi(k, t) \propto k\left(1-\frac{k}{C}\right)
\label{verhulst}
\end{align}
be, at time $t$, the probability of a new incoming vertex linking an old one that has degree $k$. Apart from the preferential linking term that is proportional to $k$, the term $\left(1-k/C\right)$ decreases $\Pi$ with $k$ and also establishes an upper bound for the degree, which is $C$, and is called the carrying capacity. This establishes also an upper bound for the number of distinct degrees (a recent study on this topic can be found in  \cite{KR13}). As will be seen later, this model presents some properties that forbid a complete analytical treatment. Note that in the present work, the condition $C>2$ will always be assumed, because the initial condition requires, at least, $C=3$ to allow the connection with a third node. One can intuitively notice that if one adopts a value of $C$ not so large, the network will display a \textquotedblleft chain-like\textquotedblright structure.

 %%%%%%%%%%%%%%%%%%%%%%%%%%%%%%%%%%%%%%%%%%%%%%%%%%%%%%%%%%%%%%%%%%%%%%%%%%%%%%%%%%%%%%%%%%%%%%%%%%%%%%%%%%%%%%
 %%%%%%%%%%%%%%%%%%%%%%%%%%%%%%%%%%%%%%%%%%%%%%%%%%%%%%%%%%%%%%%%%%%%%%%%%%%%%%%%%%%%%%%%%%%%%%%%%%%%%%%%%%%%%%
 %%%%%%%%%%%%%%%%%%%%%%%%%%%%%%%%%%%%%%%%%%%%%%%%%%%%%%%%%%%%%%%%%%%%%%%%%%%%%%%%%%%%%%%%%%%%%%%%%%%%%%%%%%%%%%

\section{Model}

The form of linking probability given by (\ref{verhulst}) follows the idea of Malthus-Verhulst model for population dynamics. The master equation of the model is then
\begin{align}
p(k, s, t+1) = \frac{k-1}{D(t)}\left(1-\frac{k-1}{C}\right)p(k-1, s, t) + \left[1-\frac{k}{D(t)}\left(1-\frac{k}{C}\right)\right]p(k, s, t)\,,
\label{me_p}
\end{align}
where
\begin{align}
D(t) := \sum_{s=1}^{t}k(s,t)\left(1-\frac{k(s,t)}{C}\right)
\label{norm}
\end{align}
is the normalisation factor, and it is essential to examine both stationary and time-dependent regimes. The function $k(s,t)$ is the degree of vertex $s$ at time $t$, and since the mean degree is $\overline{k}(t):=\sum_{s=1}^{t}k(s,t)/t=2$ for any time, a relation between $D(t)$ and the second moment, $\overline{k^{2}}(t):=\sum_{s=1}^{t}k^{2}(s,t)/t$, arises:
\begin{align}
\frac{D(t)}{t} = 2 - \frac{\overline{k^{2}}(t)}{C}\,.
\label{Nk2}
\end{align}
Note that both $D(t)$ and $\overline{k^{2}}(t)$ depend on the particular realisation of the graph (this fact will not be indicated explicitly in the notation), but it is expected that the limit
\begin{align}
\alpha_{\infty} := \lim_{t\rightarrow\infty}\alpha(t)\,,\;\textnormal{ where }\;\alpha(t):=\frac{D(t)}{t}\,,
\label{alpha}
\end{align}
should exist, since the second moment is always finite due to the carrying capacity, and is unique.

Finally, the master equation for the time-dependent degree distribution, 
\begin{align}
P(k, t) := \frac{1}{t}\sum_{s=1}^{t}p(k, s, t)\,,
\label{defPkt}
\end{align}
 is cast as
\begin{align}
\left(t+1\right)P(k, t+1) = tP(k, t) + \frac{t}{D(t)}F(k-1)P(k-1, t) - \frac{t}{D(t)}F(k)P(k, t) + p(k, s=t+1, t+1)\,,
\label{mePkt}
\end{align}
where
\begin{align}
F(k) := k\left(1-\frac{k}{C}\right)\,.
\label{F}
\end{align}
The boundary condition is $p(k, s=t+1, t+1)=\delta_{k,1}$, as seen before. The relation (\ref{Nk2}) can also be derived from this master equation.

 %%%%%%%%%%%%%%%%%%%%%%%%%%%%%%%%%%%%%%%%%%%%%%%%%%%%%%%%%%%%%%%%%%%%%%%%%%%%%%%%%%%%%%%%%%%%%%%%%%%%%%%%%%%%%%
 %%%%%%%%%%%%%%%%%%%%%%%%%%%%%%%%%%%%%%%%%%%%%%%%%%%%%%%%%%%%%%%%%%%%%%%%%%%%%%%%%%%%%%%%%%%%%%%%%%%%%%%%%%%%%%
 %%%%%%%%%%%%%%%%%%%%%%%%%%%%%%%%%%%%%%%%%%%%%%%%%%%%%%%%%%%%%%%%%%%%%%%%%%%%%%%%%%%%%%%%%%%%%%%%%%%%%%%%%%%%%%

\section{Stationary regime}

\begin{center}
\begin{figure}
\epsfig{file=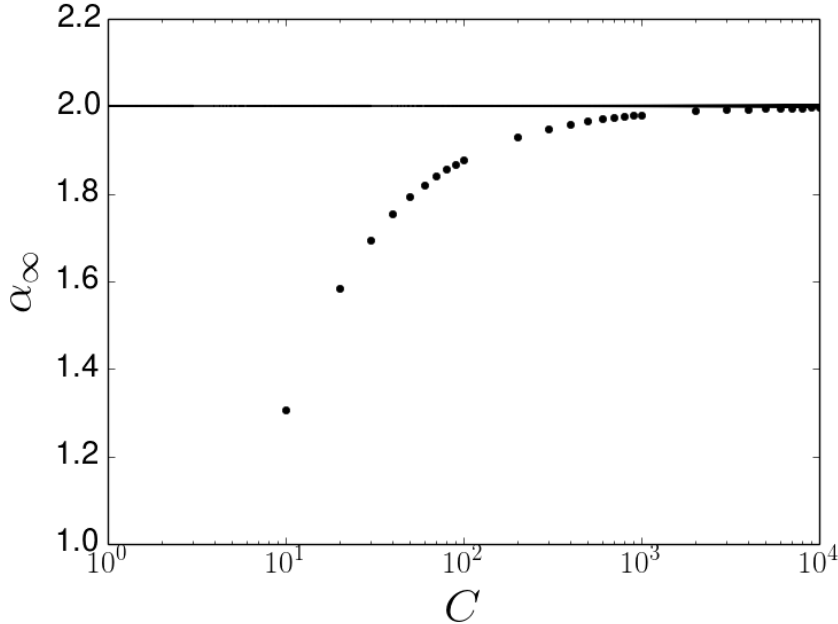, scale=0.6}
\caption{Values of $\alpha_{\infty}$ as a function of carrying capacity $C$. These values were obtained numerically from equations (\ref{pk_stat}) and (\ref{alpha_k2_infty}).}
\label{fig1}
\end{figure}
\end{center}

Assuming the existence of $P(k):=\lim_{t\rightarrow\infty}P(k, t)$, the master equation (\ref{mePkt}) leads to a recursive relation in the stationary regime,
\begin{align}
\big[\alpha_{\infty}+F(k)\big]P(k) = F(k-1)P(k-1)+\alpha_{\infty}\delta_{k,1}\,.
\label{pk_stat}
\end{align}
The degree distribution is, then,
\begin{align}
P(k) = \frac{\alpha_{\infty}\displaystyle\prod_{n=1}^{k-1}F(n)}{\displaystyle\prod_{n=1}^{k}\big[ \alpha_{\infty} + F(n) \big]}  = \frac{C!\left(k-1\right)!}{\left(C-k\right)!}\frac{\Gamma\left(\frac{Cb_{-}}{2}\right)}{\Gamma\left(\frac{Cb_{+}}{2}\right)}\frac{\Gamma\left(\frac{Cb_{+}}{2}-k\right)}{\Gamma\left(k+\frac{Cb_{-}}{2}+1\right)}\,,
\label{sol_stat}
\end{align}
where
\begin{align}
b_{\pm}:=\sqrt{1+\frac{4\alpha_{\infty}}{C}}\pm 1\,,
\end{align}
and $\alpha_{\infty}$ can be determined numerically by the bisection method through the equations
\begin{align}
\alpha_{\infty} = 2 - \frac{1}{C}\sum_{k=1}^{C}k^{2}P(k)
\label{alpha_k2_infty}
\end{align}
and (\ref{pk_stat}). The value of $\alpha_{\infty}$ depends on the size of the carrying capacity, as shown in figure \ref{fig1}. As $C$ increases, its value approaches $2$, which is the Barab\'asi-Albert limit.

\begin{center}
\begin{figure}
\epsfig{file=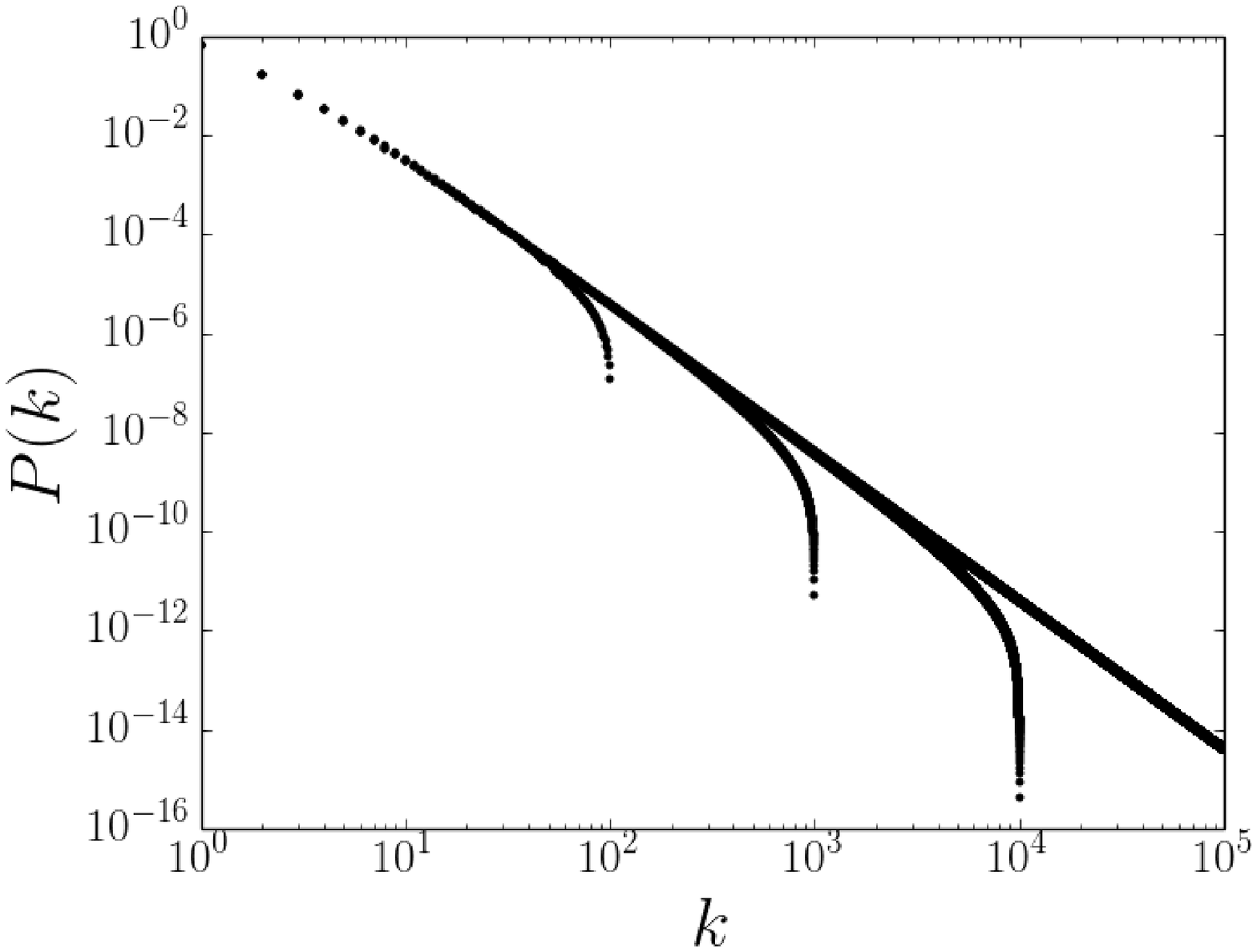, scale=0.6}
\caption{Static degree distribution obtained from equation (\ref{pk_stat}) after determining $\alpha_{\infty}$. From left to right: $C=10^{2}, 10^{3}, 10^{4}$ and $C=\infty$ (Barab\'asi-Albert limit).}
\label{fig2}
\end{figure}
\end{center}

The exact value of $\alpha_{\infty}$ allows one to determine the stationary degree distribution, as shown in figure \ref{fig2}.

The distribution for large degrees can be estimated from (\ref{pk_stat}) by converting it into a differential equation. This procedure leads to
\begin{align}
P(k) \sim \frac{1}{k^{\alpha_{\infty}+1}}\left(1-\frac{k}{C}\right)^{\alpha_{\infty}-1}\,,
\label{P_1<k<C}
\end{align}
which is not a simple power-law (and even the \textquotedblleft power-law term\textquotedblright does not behave as $k^{-3}$ like in Barab\'asi-Albert case), but it contains a term that decreases the distribution as $k$ approaches the maximum capacity $C$.

 %%%%%%%%%%%%%%%%%%%%%%%%%%%%%%%%%%%%%%%%%%%%%%%%%%%%%%%%%%%%%%%%%%%%%%%%%%%%%%%%%%%%%%%%%%%%%%%%%%%%%%%%%%%%%%
 %%%%%%%%%%%%%%%%%%%%%%%%%%%%%%%%%%%%%%%%%%%%%%%%%%%%%%%%%%%%%%%%%%%%%%%%%%%%%%%%%%%%%%%%%%%%%%%%%%%%%%%%%%%%%%
 %%%%%%%%%%%%%%%%%%%%%%%%%%%%%%%%%%%%%%%%%%%%%%%%%%%%%%%%%%%%%%%%%%%%%%%%%%%%%%%%%%%%%%%%%%%%%%%%%%%%%%%%%%%%%%

\section{Time-dependent regime and accumulation of hubs}

The master equation (\ref{mePkt}) can be treated numerically by generating graphs according to linking probability $\Pi(k,t)=F(k)/D(t)$. One can compute the normalisation factor and degree distribution associated to this particular graph, and the average over all generated graphs leads to the degree distribution $P(k,t)$ of the system. The error bars turned out to be small, and they could not be represented in the figures below. For a fixed time $t$, the degree distribution has typically the form shown in figure \ref{fig3} below, and it has two stages in its time evolution.

In the first stage, when $t<C$, the value of $P(k,t)$ is zero for any degree numerically larger than $t$, and $P(k=t,t)$ is the non-zero minimum. Therefore, as time passes, the distribution $P(k,t)$ acquires more non-zero values as a function of $k$. This process continues until the time reaches $t=C$, and $P(k=C,t=C)$ is the least non-zero minimum that the function $P$ can achieve.

In the second stage, when $t>C$, the vertices can have, in principle, the maximum value allowed, which is $C$. In this stage, there is a rearrangement of the curve $P(k,t)$ until it reaches the stationary form in $t\rightarrow\infty$. If $t>C$, all vertices can be \textquotedblleft degenerated\textquotedblright, and this property is reflected in the increase of the curve $P(k,t)$ for high values of $k$.

\begin{figure}
\epsfig{file=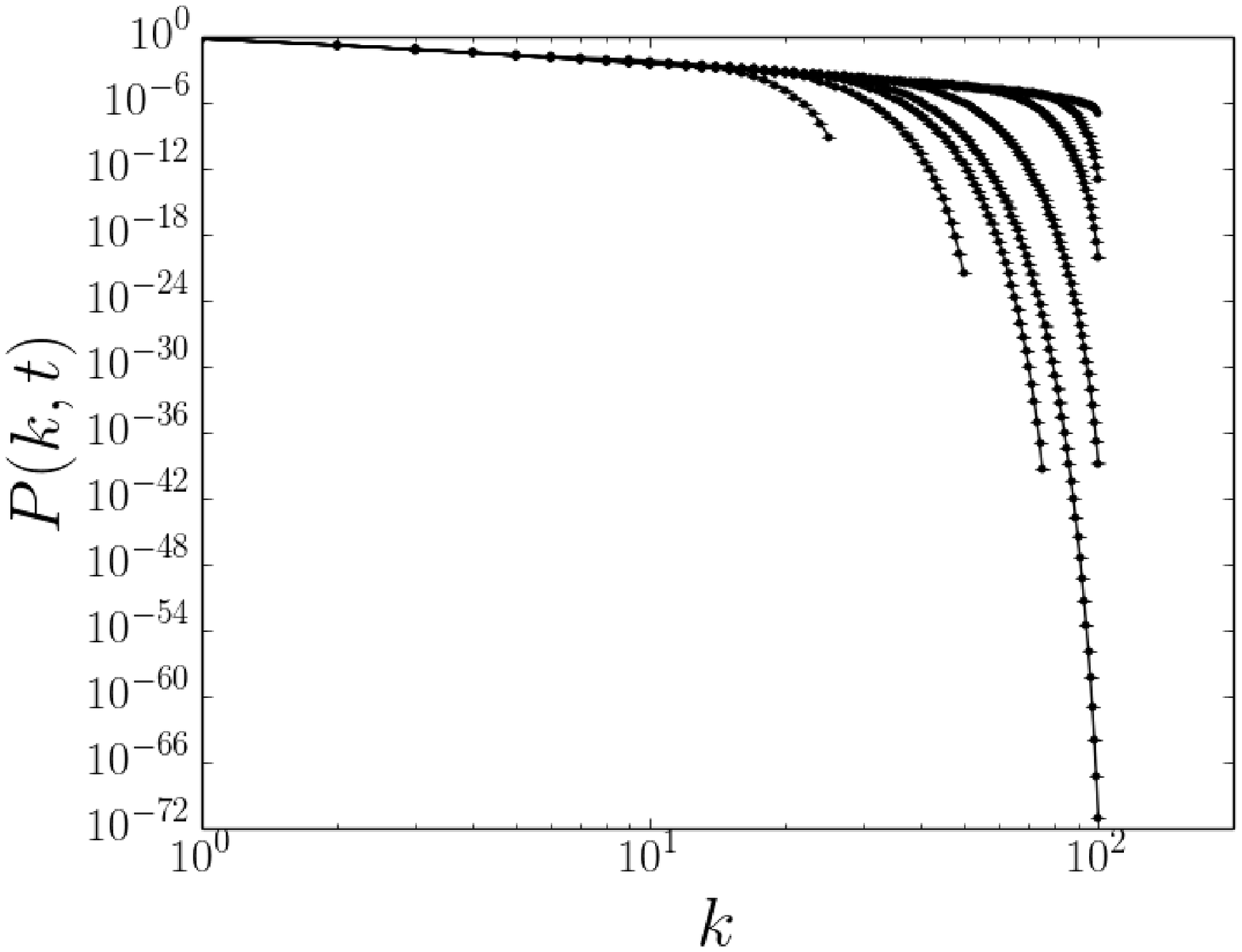, scale=0.43}
\epsfig{file=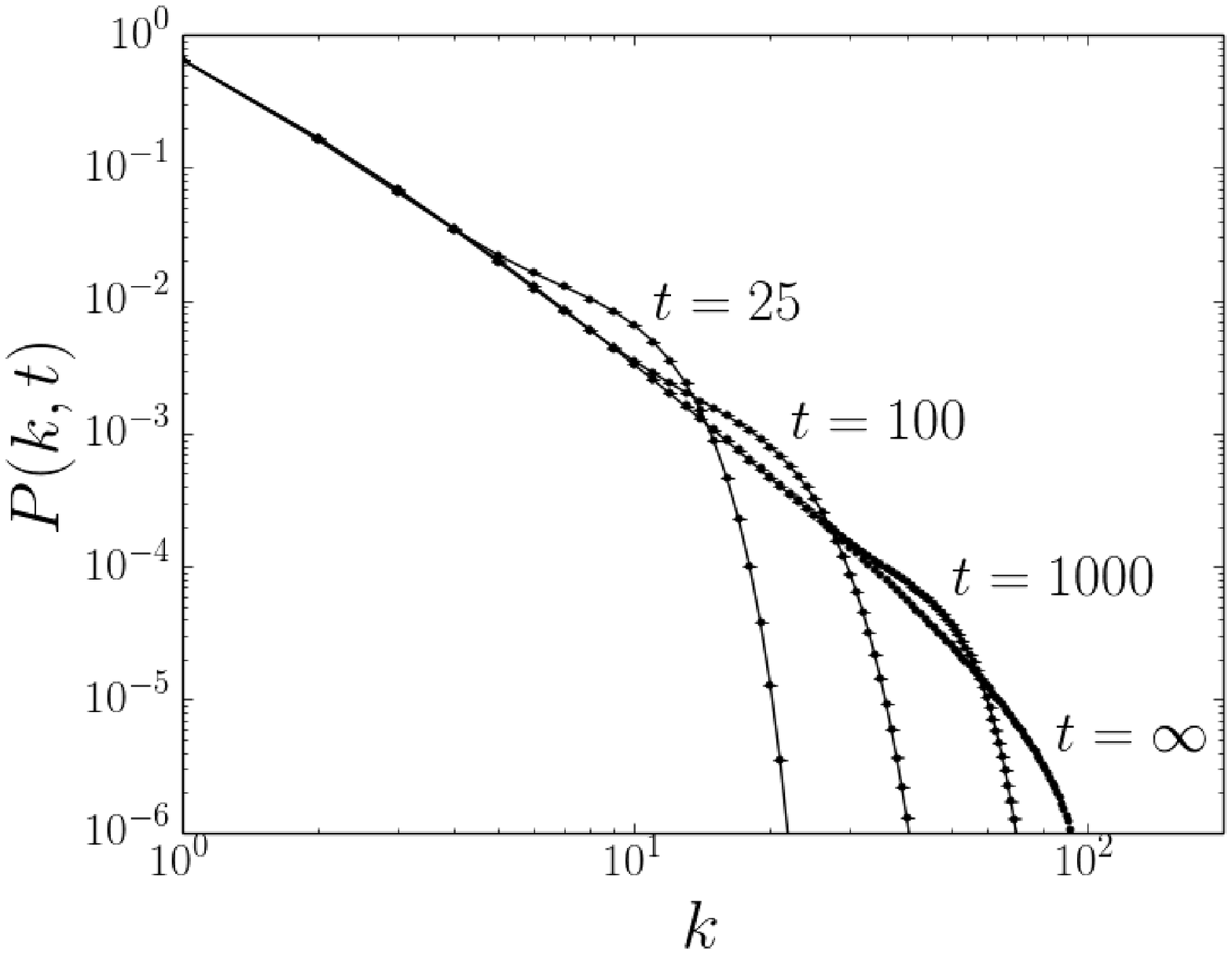, scale=0.43}
\caption{Time-dependent degree distribution for $C=100$. (a) From left to right: $t=25$, $50$, $75$, $100$, $200$, $1000$, $5000$ (curves obtained from $10^{4}$ realisations of graphs) and $t=\infty$ (stationary degree distribution -- obtained numerically in the previous section). (b) The curves for $t=25$, $100$, $1000$ and $t=\infty$ at a restricted region. For finite times, a \textquotedblleft bump\textquotedblright in the time-dependent degree distribution appears.}
\label{fig3}
\end{figure}

In figure \ref{fig3}b, one can observe a phenomenon which is not present in many of the previous models with bounded degree effects, but found in \cite{KR02}. For finite times, the degree distribution (as a function of degree $k$) exhibits a \textquotedblleft bump\textquotedblright\, before its fall because of the finiteness of carrying capacity: the degree distribution $P(k,t)$, for a fixed time, deviates to larger values (when compared to the stationary distribution $P(k)$) before it decreases. This occurs as an effect of two opposite behaviours of the linking probability $\Pi$. Due to the attractive effect, nodes with many connections attract new ones, and a concentration of vertices with large degree (the hubs) emerges. This is a temporary \textquotedblleft saturation\textquotedblright\, of hubs that accumulated links and become unattractive to receive more connections. The region of this concentration in the graph moves to values closer to the maximum degree $C$ as time passes. In the end, this \textquotedblleft bump\textquotedblright disappears. Note that this phenomenon realizes as a consequence of a competition between an attractive term (the linear preferential linking) and repulsive one (the Verhulst correction), and cannot be predicted by imposing an abrupt cutoff in degree. The next section will consider a toy model where the linking probability is based on repulsive term only. The idea is to show that the attractive term is also essential to display the \textquotedblleft bump\textquotedblright\, in the degree distribution.

\begin{figure}
\epsfig{file=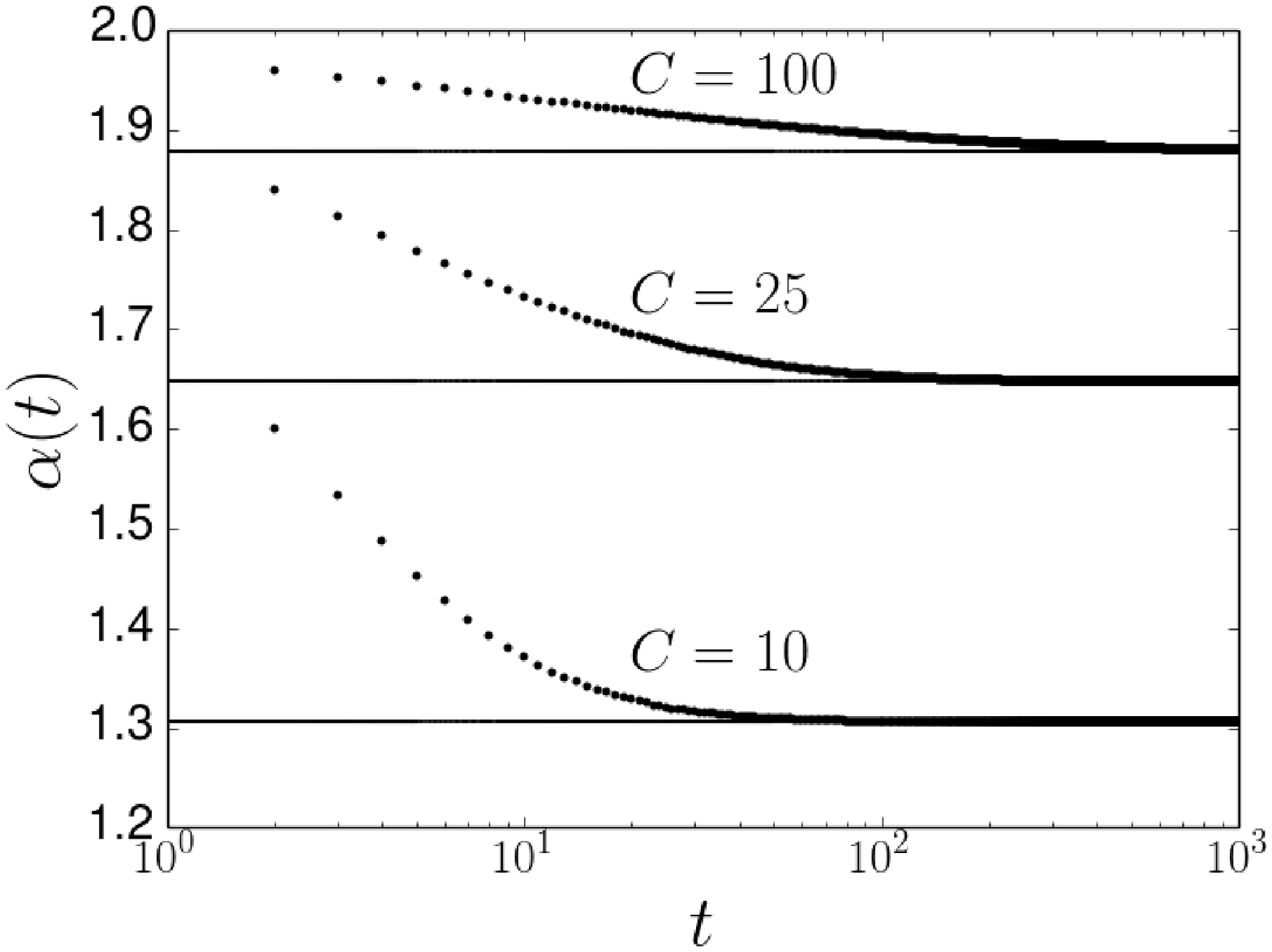, scale=0.43}
\caption{Dependence of $\alpha(t)$ on time obtained from $10^{4}$ realisations of the graphs (see section 3) for different values of the carrying capacity $C$ (in the Barab\'asi-Albert limit, the curve is a horizontal line $\alpha(t)=\alpha_{\infty}=2$). The horizontal lines are the stationary values $\alpha_{\infty}$ (see section 4), corresponding to each value of $C$; from top to bottom: $\alpha_{\infty}=1.8787\ldots$, $1.6481\ldots$ and $1.3055\ldots$ for $C=100$, $25$ and $10$, respectively.}
\label{alphaxt}
\end{figure}

\begin{figure}
\epsfig{file=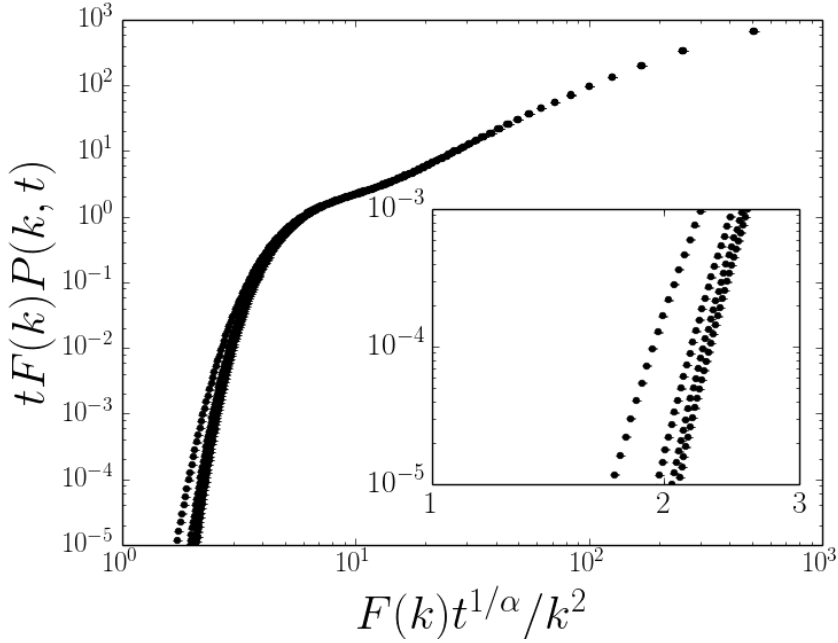, scale=0.6}
\caption{Scaling behaviour of time-dependent degree distribution. In the bottom right part, the same graph in extended scale. The curves from left to right correspond to $C=250, 500, 750$ and $1000$ ($t=1000$). The error bars are smaller than the size of points.}
\label{fig4}
\end{figure}

In the large time asymptotic regime, the probability $p(k,s,t)$ has a scaling $p(k,s,t)\sim p(k,s/t)$, as shown explicitly in the Appendix. Furthermore, assuming that the ratio $D(t)/t$ is $\alpha_{\infty}$ for large times (figure \ref{alphaxt} shows the convergence of $\alpha(t)$ toward $\alpha_{\infty}$ with time for some values of the carrying capacity), and denoting $r:=s/t$, the continuum approximation leads the master equation (\ref{me}) to
\begin{align}
\alpha_{\infty} r\frac{\partial}{\partial r}p(k,r) = \frac{\partial}{\partial k}\left[ F(k)p(k,r)\right]\,,
\label{kr}
\end{align}
which can be solved as
\begin{align}
p(k,r)=\frac{1}{F(k)}\psi\left(\frac{F(k)}{k^{2}}\frac{1}{r^{\frac{1}{\alpha_{\infty}}}}\right)\,,
\end{align}
where $\psi$ is a differentiable function. This approach can also be invoked for the master equation (\ref{mePkt}), which yields the partial differential equation
\begin{align}
\alpha_{\infty}\frac{\partial}{\partial t}\big[tP(k,t)\big] = -\frac{\partial}{\partial k}\big[F(k)P(k,t)\big]
\label{pde}
\end{align}
for the time-dependent degree distribution. The solution
\begin{align}
P(k, t) = \frac{1}{tF(k)}\varphi\left(\frac{F(k)}{k^{2}}t^{\frac{1}{\alpha_{\infty}}}\right)\,,
\label{cont_Pkt}
\end{align}
where $\varphi$ is a differentiable function, and from the stationary condition for $k\gg 1$, it follows from (\ref{P_1<k<C}) that
\begin{align}
\varphi(x) \sim x^{\alpha_{\infty}}
\label{varphi}
\end{align}
for $x\gg 1$. In figure \ref{fig4}, the form of $\varphi$ is shown for some values of $C$, and the hypothesis of assuming $\alpha(t)\approx\alpha_{\infty}$ is tested: one sees that the curves collapse into a single one, except for low values of $F(k)t^{1/\alpha_{\infty}}/k^{2}$. When this happens, it means that $k$ and/or $t$ are not high enough, violating the conditions necessary to derive (\ref{cont_Pkt}).

 %%%%%%%%%%%%%%%%%%%%%%%%%%%%%%%%%%%%%%%%%%%%%%%%%%%%%%%%%%%%%%%%%%%%%%%%%%%%%%%%%%%%%%%%%%%%%%%%%%%%%%%%%%%%%%
 %%%%%%%%%%%%%%%%%%%%%%%%%%%%%%%%%%%%%%%%%%%%%%%%%%%%%%%%%%%%%%%%%%%%%%%%%%%%%%%%%%%%%%%%%%%%%%%%%%%%%%%%%%%%%%
 %%%%%%%%%%%%%%%%%%%%%%%%%%%%%%%%%%%%%%%%%%%%%%%%%%%%%%%%%%%%%%%%%%%%%%%%%%%%%%%%%%%%%%%%%%%%%%%%%%%%%%%%%%%%%%

\section{A simplified toy model}

\begin{center}
\begin{figure}
\epsfig{file=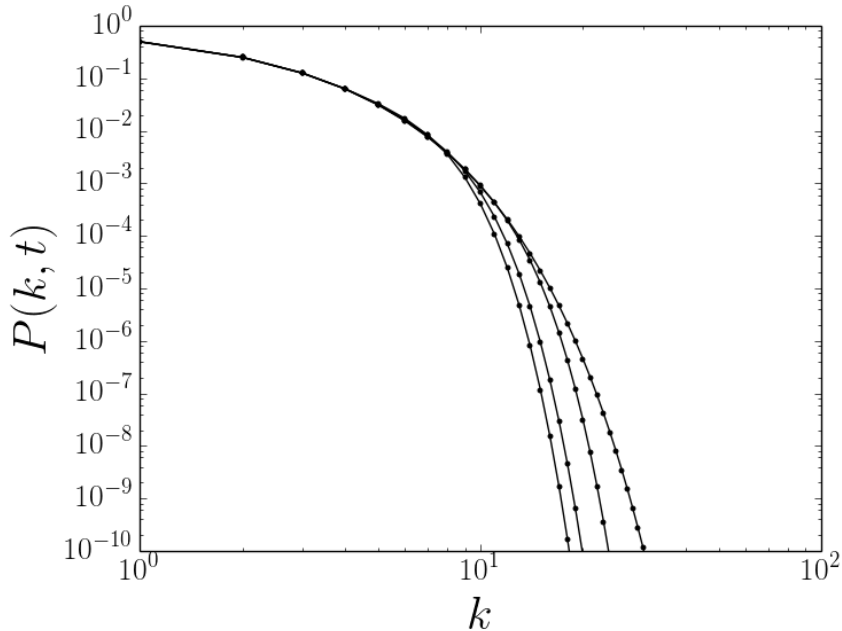, scale=0.6}
\caption{Time-dependent degree distribution for the toy model with $C=100$. From left to right, $t=50, 100, 1000$ and $t=\infty$ (stationary solution).}
\label{fig5}
\end{figure}
\end{center}

The concentration of hubs observed in the previous section will be searched in a model where the (linear) preferential linking term is absent. Instead of studying a network dynamics that obeys (\ref{verhulst}), consider the conditional probability
\begin{align}
\Pi_{0} = \Pi_{0}(k, t) \propto \left(1-\frac{k}{C}\right)
\label{prePi0}
\end{align}
as the rule that determines how the system evolves. The other conditions (initial and boundary conditions) are taken to be the same as the model discussed in the previous section. In this simplified model, a new vertex is not attracted to a popular one, but it takes into account the repulsive factor only. The normalisation factor of (\ref{prePi0}) is easily found, and the exact expression for the linking probability is
\begin{align}
\Pi_{0}(k, t) = \frac{1-k/C}{t\left(1-2/C\right)}\,.
\label{Pi0}
\end{align}
Therefore, from the master equation (\ref{me}), one can calculate the mean degree of vertex $s$ at time $t$,
\begin{align}
\overline{k}(s, t) := \sum_{k=1}^{C}kp(k, s, t) = \sum_{n=0}^{t-s-1}\rho(t-n-1)\prod_{m=1}^{n}\left[1-\frac{\rho(t-m)}{C}\right] + \prod_{m=1}^{t-s}\left[1-\frac{\rho(t-m)}{C}\right]\,,
\label{kst0}
\end{align}
where one defines $\prod_{n=1}^{0}\left(\cdots\right)\equiv 1$ and
\begin{align}
\rho(t) := \frac{1}{t\left(1-2/C\right)}\,.
\label{rho}
\end{align}
In the asymptotic limit of large times, it is possible to show that
\begin{align}
\overline{k}(s, t) \sim C - \left(C-1\right)\left(\frac{s}{t}\right)^{\frac{1}{C-2}}\,,
\label{k0_asymp}
\end{align}
which means that the mean degree of a vertex $s$ behaves as a function of $s/t$ for $t\gg 1$. From the master equation (\ref{me}), one can evaluate the time-dependent degree distribution numerically, as shown in figure \ref{fig5}. The concentration of higher degrees, which was responsible for a \textquotedblleft bump\textquotedblright\, in the figure \ref{fig3}b, is not seen in this simplified model. This means that the repulsive term $\left(1-k/C\right)$ alone in the linking probability $\Pi$ is not sufficient to display this behaviour.

The static degree distribution, which can be obtained from $P(k):=\lim_{t\rightarrow\infty}P(k, t)$, is analytically accessible, and it is equal to
\begin{align}
P(k) = \frac{C-2}{2C-3}\frac{\left(C-1\right)!}{\left(2C-4\right)!}\frac{\left(2C-k-3\right)!}{\left(C-k\right)!}\,.
\label{Pk0}
\end{align}

Finally, the probability $p(k, s, t)$ in this model can be evaluated exactly in the limit of large times,
\begin{align}
p(k, s, t) \sim \frac{1}{\left(k-1\right)!}\frac{\left(C-1\right)!}{\left(C-k\right)!}\left(\frac{s}{t}\right)^{\frac{C-k}{C-2}}\left[1-\left(\frac{s}{t}\right)^{\frac{1}{C-2}}\right]^{k-1}\,,
\label{pkst0}
\end{align}
and it is a function of $s/t$.

 %%%%%%%%%%%%%%%%%%%%%%%%%%%%%%%%%%%%%%%%%%%%%%%%%%%%%%%%%%%%%%%%%%%%%%%%%%%%%%%%%%%%%%%%%%%%%%%%%%%%%%%%%%%%%%
 %%%%%%%%%%%%%%%%%%%%%%%%%%%%%%%%%%%%%%%%%%%%%%%%%%%%%%%%%%%%%%%%%%%%%%%%%%%%%%%%%%%%%%%%%%%%%%%%%%%%%%%%%%%%%%
 %%%%%%%%%%%%%%%%%%%%%%%%%%%%%%%%%%%%%%%%%%%%%%%%%%%%%%%%%%%%%%%%%%%%%%%%%%%%%%%%%%%%%%%%%%%%%%%%%%%%%%%%%%%%%%

\section{Conclusions}

The Barab\'asi-Albert model is a growing network that displays a power-law behaviour in degree distribution. Nevertheless, it allows an indefinite growth of degree, and this is not present in many realistic systems. In this work, a scheme where this undesirable property is suppressed was proposed by introducing a carrying capacity $C$, following the idea of Verhulst in the context of population growth. Both stationary and time-dependent regime of this proposed model was analysed, and a clear modification in degree distribution is observed when the degree approaches $C$. Moreover, there is a close connection between the second moment of the degree and the normalisation of the linking probability, and some other analytical results for asymptotically large times and degrees were established. In the limit $C\rightarrow\infty$, all the known results from Barab\'asi-Albert were recovered.

In the transient case, the behaviour of degree distribution, as a function of degree, was characterised for fixed times. Furthermore, a scaling form for the degree distribution was found.

In the present model, a concentration of hubs was observed for finite times. This phenomenon is evident from the analysis of the time-dependent degree distribution $P(k,t)$, which displays a \textquotedblleft bump\textquotedblright for a fixed time. This behaviour is attributed to the competition between the usual BA-like attractive term and a Verhulst-like repulsive one, as shown in section 5. One should remark that this phenomenon emerged because of the \textquotedblleft smoothness\textquotedblright of the repulsive term, which made the system \textquotedblleft be aware\textquotedblright of the finiteness before the vertices achieve the maximum allowed number of connections. If the bound of degree is introduced through a cutoff \cite{ASBS00}, this behaviour is not displayed.

The mechanism of degree finiteness is introduced by the Verhulst scheme (\ref{verhulst}), which is a product of a linear preferential term and a repulsive term $\left(1-k/C\right)$. One may study the case where nonlinear functions are chosen instead of linear ones. In particular, the persistence of the accumulation of hubs (the \textquotedblleft bump\textquotedblright of figure 3b), which is a trademark of the awareness of degree finiteness by the system, can be studied for different types of linking probabilities. It is known that the inclusion of nonlinearity can lead to new behaviours, as in the case of BA model, where nonlinear preferential linking probability destroys the scale-free profile of the degree distribution \cite{KR01, KRL00}.

 %%%%%%%%%%%%%%%%%%%%%%%%%%%%%%%%%%%%%%%%%%%%%%%%%%%%%%%%%%%%%%%%%%%%%%%%%%%%%%%%%%%%%%%%%%%%%%%%%%%%%%%%%%%%%%
 %%%%%%%%%%%%%%%%%%%%%%%%%%%%%%%%%%%%%%%%%%%%%%%%%%%%%%%%%%%%%%%%%%%%%%%%%%%%%%%%%%%%%%%%%%%%%%%%%%%%%%%%%%%%%%
 %%%%%%%%%%%%%%%%%%%%%%%%%%%%%%%%%%%%%%%%%%%%%%%%%%%%%%%%%%%%%%%%%%%%%%%%%%%%%%%%%%%%%%%%%%%%%%%%%%%%%%%%%%%%%%

\section{Acknowledgements}

MOH is thankful to D. B. Liarte and M. J. de Oliveira for useful conversations and to the Instituto de F\'isica da Universidade de S\~ao Paulo, where he had access to many computational facilities. HLCG thanks PNPD-CAPES (Ed. 82/2014) for financial support.

 %%%%%%%%%%%%%%%%%%%%%%%%%%%%%%%%%%%%%%%%%%%%%%%%%%%%%%%%%%%%%%%%%%%%%%%%%%%%%%%%%%%%%%%%%%%%%%%%%%%%%%%%%%%%%%
 %%%%%%%%%%%%%%%%%%%%%%%%%%%%%%%%%%%%%%%%%%%%%%%%%%%%%%%%%%%%%%%%%%%%%%%%%%%%%%%%%%%%%%%%%%%%%%%%%%%%%%%%%%%%%%
 %%%%%%%%%%%%%%%%%%%%%%%%%%%%%%%%%%%%%%%%%%%%%%%%%%%%%%%%%%%%%%%%%%%%%%%%%%%%%%%%%%%%%%%%%%%%%%%%%%%%%%%%%%%%%%

\section{Appendix - Asymptotic form of $p(k, s, t)$}

\begin{center}
\begin{figure}
\epsfig{file=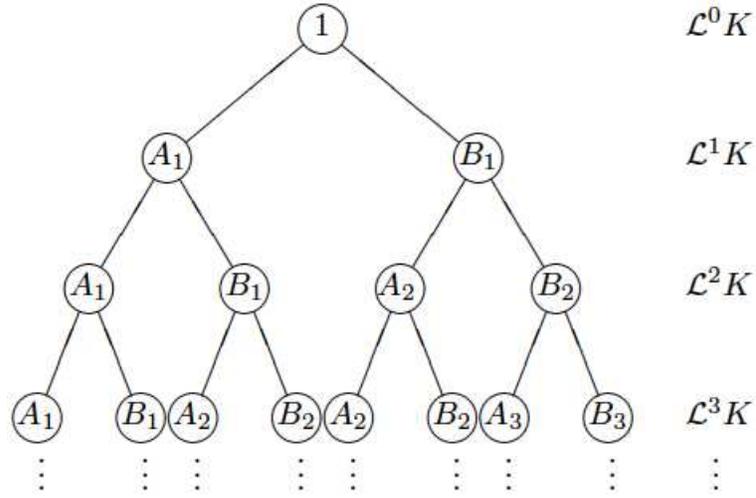, scale=0.6}
\caption{Coefficients of $\mathcal{L}^{n}K$.}
\label{fig6}
\end{figure}
\end{center}

In this appendix, it will be shown that in asymptotic regime ($t\gg 1$), the probability $p(k,s,t)$ behaves as $p(k,s/t)$. From the master equation (\ref{me_p}), and using $N(t\gg 1)\sim\alpha_{\infty} t$, one has
\begin{align}
t\frac{\partial}{\partial t}p(k, s, t) = \frac{k-1}{\alpha_{\infty}}\left(1-\frac{k-1}{C}\right)p(k-1, s, t) - \frac{k}{\alpha_{\infty}}\left(1-\frac{k}{C}\right)p(k, s, t)\,.
\label{ap1}
\end{align}
Introducing the Z-transform,
\begin{align}
p^{Z}(K,s,t) = \sum_{k}K^{k}p(k,s,t)\,,
\end{align}
the master equation (\ref{ap1}) can be cast as
\begin{align}
\nonumber t\frac{\partial}{\partial t}p^{Z}(K, s, t) &= -\frac{1}{\alpha_{\infty} C}\left(C-1\right)K\left(1-K\right)\frac{\partial}{\partial K}p^{Z}(K, s, t) + \frac{1}{\alpha_{\infty} C}K^{2}\left(1-K\right)\frac{\partial^{2}}{\partial K^{2}}p^{Z}(K, s, t) \\
 & = -\frac{C}{\alpha_{\infty}}\mathcal{L}p^{Z}(K, s, t)\,,
\label{ap2}
\end{align}
where $\mathcal{L}:=\left(\frac{C-1}{C^{2}}\right)K\left(1-K\right)\frac{\partial}{\partial K}-\left(\frac{K}{C}\right)^{2}\left(1-K\right)\frac{\partial^{2}}{\partial K^{2}}$ is a differential operator. The differential equation (\ref{ap2}) can be solved formally as
\begin{align}
p^{Z}(K, s, t) = t^{-\frac{C}{\alpha_{\infty}}\mathcal{L}}f(K, s)\,,
\end{align}
where $f$ is a function that can be determined from the boundary condition $p^{Z}(K,s,t=s)=K$. This implies
\begin{align}
p^{Z}(K, s, t) = \left(\frac{s}{t}\right)^{\frac{C}{\alpha_{\infty}}\mathcal{L}}K = \sum_{n=0}^{\infty}\frac{1}{n!}\left[\frac{C}{\alpha_{\infty}}\ln\left(\frac{s}{t}\right)\right]^{n}\mathcal{L}^{n}K\,,
\end{align}
and one can already see that $p(k,s,t)$ (which is the inverse Z-transform of $p^{Z}(K,s,t)$) is a function of $p(k,s/t)$. The remaining part of this appendix will search for a form of $p(k,s,t)$ that does not contain the operator $\mathcal{L}$ explicitly (this does not mean that the resulting formula is simple). Note that
\begin{align}
\mathcal{L} K^{m} = A_{m}K^{m} + B_{m}K^{m+1}\,,
\label{LKm}
\end{align}
with
\begin{align}
A_{m}:= \frac{m}{C}\left(1-\frac{m}{C}\right) =: -B_{m}\;\;\textrm{for}\;\; m\geq 1\;\textrm{ and }\; B_{0}=1\,.
\end{align}
Using this relation repeatedly, one sees that $\mathcal{L}^{n}K$ is a polynomial, and the diagram of figure \ref{fig6} yields an algorithm for its coefficients.

Schematically, one has $A_{m}\rightarrow (A_{m}, B_{m})$ and $B_{m}\rightarrow (A_{m+1}, B_{m+1})$ when passing from $\mathcal{L}^{m}K$ to $\mathcal{L}^{m+1}K$. Moreover, $A_{m}$ and $B_{m}$ are coefficients of $K^{m}$ and $K^{m+1}$, respectively. Therefore,
\begin{widetext}
\begin{align}
\nonumber \mathcal{L}^{n} K & = \sum_{r=0}^{n}\sum_{{i_{1},\cdots,i_{r+1}\atop i_{1}+\cdots +i_{r+1}=n-r}}\left(A_{1}^{i_{1}}B_{1}\right)\left(A_{2}^{i_{2}}B_{2}\right)\cdots\left(A_{r}^{i_{r}}B_{r}\right)A_{r+1}^{i_{r+1}}K^{r+1} \\
\nonumber & = \sum_{r=0}^{n}\sum_{i_{1},\cdots,i_{r+1}=0}^{\infty}\left(A_{1}^{i_{1}}B_{1}\right)\left(A_{2}^{i_{2}}B_{2}\right)\cdots\left(A_{r}^{i_{r}}B_{r}\right)A_{r+1}^{i_{r+1}}K^{r+1}\delta_{i_{1}+\cdots+i_{r+1}, n-r} \\
 & = \sum_{r=0}^{n}\frac{1}{\left(n-r\right)!}B_{1}\cdots B_{r}\,\frac{d^{n-r}}{dz^{n-r}}\left[\left(1-A_{1}z\right)^{-1}\cdots\left(1-A_{r+1}z\right)^{-1}\right]\Bigg|_{z\downarrow 0}K^{r+1}\,,
\end{align}
\end{widetext}
where the integral representation of Kronecker delta was invoked above. Then, from
\begin{align}
p^{Z}(K, s, t) = \sum_{n=0}^{\infty}\frac{1}{n!}\left[\frac{C}{\alpha_{\infty}}\ln\left(\frac{s}{t}\right)\right]^{n} \sum_{r=0}^{n}\frac{1}{\left(n-r\right)!}B_{1}\cdots B_{r}\frac{d^{n-r}}{dz^{n-r}}\left[\left(1-A_{1}z\right)^{-1}\cdots\left(1-A_{r+1}z\right)^{-1}\right]\Bigg|_{z\downarrow 0}K^{r+1}\,,
\end{align}
and since
\begin{align}
B_{1}\cdots B_{k-1} & = \frac{\left(-1\right)^{k-1}}{C^{2\left(k-1\right)}}\frac{\left(k-1\right)!\left(C-1\right)!}{\left(C-k\right)!}\,,
\end{align}
one has finally
\begin{widetext}
\begin{align}
\nonumber p(k, s, t) & = \sum_{n=k-1}^{\infty}\frac{1}{n!}\left[\frac{C}{\alpha_{\infty}}\ln\left(\frac{s}{t}\right)\right]^{n} \frac{1}{\left(n-k+1\right)!}B_{1}\cdots B_{k-1}\,\frac{d^{n-k+1}}{dz^{n-k+1}}\left[\left(1-A_{1}z\right)^{-1}\cdots\left(1-A_{k}z\right)^{-1}\right]\Bigg|_{z\downarrow 0} \\
\nonumber & = \frac{\left(C-1\right)!\left(k-1\right)!}{\left(2C\right)^{k-1}\left(C-k\right)!}\ln^{k-1}\left(\frac{t}{s}\right)\sum_{m=0}^{\infty}\frac{1}{m!\left(m+k-1\right)!}\left[\frac{C}{\alpha_{\infty}}\ln\left(\frac{s}{t}\right)\right]^{m}  \frac{d^{m}}{dz^{m}}\left[\left(1-A_{1}z\right)^{-1}\cdots\left(1-A_{k}z\right)^{-1}\right]\Bigg|_{z\downarrow 0}\,. \\
\end{align}
\end{widetext}

\end{document}